\documentclass[aps,prl,superscriptaddress,showkeys,showpacs,twocolumn,longbibliography,nofootinbib]{revtex4-1}
\usepackage{amsmath,mathrsfs}    % need for subequations
\usepackage{graphicx}   % need for figures
\usepackage[caption=false]{subfig}
\usepackage{color}
\usepackage[colorlinks=true,linkcolor=blue,citecolor=blue,urlcolor=blue]{hyperref}   
\usepackage{bm} % for bold caligraphic font
\usepackage[inline]{enumitem}

\usepackage{graphicx}
\usepackage{textcomp}
\usepackage{hyperref}
\usepackage[capitalise]{cleveref}
\usepackage[toc,page]{appendix}
\usepackage{xcolor}
\usepackage{tikz}

\def\st{\begin{equation}}
\def\stp{\end{equation}}

\usepackage{array}   % for \newcolumntype macro
\newcolumntype{C}{>{$}c<{$}} % math-mode version of "l" column type

\definecolor{MyDarkBlue}{RGB}{22, 79, 134}
\definecolor{MyLightRed}{RGB}{200, 37, 6}

\def\x{{\bm x}}

\def\k{{\bm k}}

\def\k{{\bm k}}
\def\F{{\mathcal H}}

\def\st{\begin{equation}}
\def\stp{\end{equation}}
\def\bg{\begin{eqnarray}}
\def\nd{\end{eqnarray}}

\def\llangle{\left\langle}
\def\rrangle{\right\rangle}
\def \bes {\begin{subequations}}
\def \ees {\end{subequations}}

\def \F

\def\chemconst{{\chizero}}

\newcommand{\chizero}{{\chi_I}}
\def\bsigma{\bar{\sigma}_{\rm eq}}
\newcommand{\tred}{{\sf t}_{\sf r} }
\newcommand{\tredini}{{\sf t}_{\sf r}^0}

\def\bsigma{\bar{\sigma}_{\rm eq}}

\newcommand{\citelongpaper}{~\cite{Florio:2025lvu}}

\newcommand{\VEV}{condensate}

\begin{document}
\title{Supercooled Goldstones at the QCD chiral phase transition}
\author{Adrien Florio}
\email[]{aflorio@physik.uni-bielefeld.de}
\affiliation{Fakultät für Physik, Universität Bielefeld, D-33615 Bielefeld, Germany}
\affiliation{Department of Physics, Brookhaven National Laboratory, Upton, New
York 11973-5000, USA}
\author{Eduardo Grossi}
\email[]{eduardo.grossi@unifi.it}
\affiliation{Dipartimento di Fisica, Universit\`a di Firenze and INFN Sezione
    di Firenze, via G. Sansone 1,
50019 Sesto Fiorentino, Italy}
\author{Aleksas Mazeliauskas}
\email[]{a.mazeliauskas@thphys.uni-heidelberg.de}
\affiliation{Institut für Theoretische Physik, Universität Heidelberg
D-69120 Heidelberg, Germany}
\author{Alexander Soloviev}
\email[]{alexander.soloviev@fmf.uni-lj.si}
\affiliation{Faculty of Mathematics and Physics, University of Ljubljana,
Jadranska ulica 19, SI-1000 Ljubljana, Slovenia}
\author{Derek Teaney}
\email[]{derek.teaney@stonybrook.edu}
\affiliation{Center for Nuclear Theory, Department of Physics and Astronomy,
Stony Brook University, New York 11794-3800, USA}

\date{\today}

\begin{abstract}
    We discuss a universal non-equilibrium enhancement of long-wavelength Goldstone bosons induced by quenches to the broken phase in Model G --- the dynamical universality class of an $O(4)$\nobreakdash-antiferromagnet and the chiral phase transition in QCD. Scaling arguments for the coarsening dynamics describing the formation of the chiral condensate predict a parametric enhancement in the infrared spectra of Goldstones, a prediction confirmed by stochastic simulations of the transition. The details of the enhancement are determined by the non-linear dynamics of a superfluid effective theory, which is a limit of Model G  reflecting the broken $O(4)$ symmetry.  
Our results translate to a parametric enhancement of low-momentum pions in heavy-ion collisions at the LHC, which are underpredicted in current hydrodynamic models without critical dynamics.
\end{abstract}

\maketitle

\textit{Introduction.} 
Experiments at the Relativistic Heavy Ion Collider (RHIC) and the Large Hadron Collider (LHC) have provided compelling
evidence that high-energy heavy-ion collisions create an extended space-time
region of Quark-Gluon Plasma (QGP)~\cite{Busza:2018rrf}. QGP is the high-temperature phase
of quantum chromodynamics (QCD) where the approximate chiral symmetry of the
QCD Lagrangian is restored~\cite{Pisarski:1983ms}. As the QGP rapidly expands
and cools, the system returns to the hadronic phase, where quarks and gluons
are confined and chiral symmetry is spontaneously broken. This transition is a
smooth crossover~\cite{Borsanyi:2018grb,HotQCD:2018pds}, but lies close in
parameter space to a second-order critical point. In the chiral limit, where
the up and down quark masses vanish, the transition is second
order~\cite{HotQCD:2019xnw,Kaczmarek:2020sif,Kotov:2021rah,Cuteri:2021ikv} and
believed to belong to the universality class of an $O(4)$ antiferromagnet. Near
this critical point, both the correlation length $\xi$ and the relaxation time
$\tau_R \propto \xi^{\zeta}$ grow, giving rise to universal dynamics. As the QCD medium expands and cools, this universal dynamics
controls  both the formation of the chiral condensate (the order parameter of the
chiral transition) and the production of soft pions (the Goldstone modes of the transition).

Many years ago the dynamical universality class of chiral QCD was identified as ``Model G'' in the classification scheme of Halperin and Hohenberg~\cite{Rajagopal:1992qz}. Early studies speculated that the rapid expansion of the QGP could lead to the formation of a disoriented chiral condensate~\cite{Bjorken:1991xr,Rajagopal:1992qz,Blaizot:1992at,Gavin:1993bs}, but these predictions were never experimentally confirmed~\cite{UA1:1982kzj,UA5:1986zif,Augusto:1999vz,MiniMax:1999mxv}. In the intervening years, hydrodynamic models for heavy-ion collisions achieved remarkable success in describing the bulk properties of these collisions, placing tight constraints on QGP transport properties~\cite{Busza:2018rrf}.
Nevertheless, a persistent tension between hydro models and increasingly precise data on the yield of low-momentum pions,  suggests that the hydrodynamic description of the QGP is incomplete~\cite{ALICE:2013mez,ALICE:2019hno,Mazeliauskas:2019ifr,Melo:2019mpn,Devetak:2019lsk,Nijs:2020roc,JETSCAPE:2020mzn,Lu:2024upk}. Moreover,  recent progress in lattice QCD has highlighted the significance of the $O(4)$ critical point to real-world QCD,  reviving interest in chiral critical dynamics and its potential phenomenological signatures~\cite{Schlichting:2019tbr,Grossi:2020ezz,Grossi:2021gqi,Cao:2022csq,Yao:2022fwm,Florio:2023kmy,Braun:2023pew,Roth:2024rbi,Roth:2024hcu}.

With this context, we study the coarsening dynamics describing the formation of the chiral condensate after a quench from the restored to the broken phase in Model G. Due to  distinctive reversible couplings (i.e. Poisson brackets)  between the $O(4)$ order parameter and QCD’s conserved charges, coarsening in this model is markedly different from other previously studied critical models such as Model A and Model B~\cite{Bray:1994zz,tauber2014critical}. Indeed, these couplings cause the Goldstone modes to behave like pseudo-particles that propagate ballistically over many correlation lengths before diffusing back into the medium. This separation of scales leads to a parametric enhancement over equilibrium of the infrared spectrum Goldstones, which persists after the quench for a parametrically long time compared to $\tau_R$. We verify these predictions through comprehensive stochastic simulations of instantaneous quenches in Model G, described in detail in our companion paper\citelongpaper. The soft pion enhancement is a rigorous prediction of QCD close to the chiral phase transition, and it is a prediction with tantalizing phenomenological implications. 

\textit{$O(4)$-antiferromagnet (Model G).} We briefly review Model G \cite{hohenberg} and refer the reader to our companion paper\citelongpaper~and previous works~\cite{Grossi:2021gqi}. Model G captures the dynamics of an $O(4)$ local order parameter $(\phi_0, \phi_1, \phi_2, \phi_3)$ coupled to partially conserved axial charges $n_A^\ell$ and conserved vector charges $ n_V^{\ell}$, which can be combined in an antisymmetric $O(4)$ tensor $n_{ab}$ with $n_{0\ell} = n_A^\ell\,$ and  $   n_{\ell_1\ell_2} =\epsilon^{\ell_1\ell_2 \ell} n_V^{\ell}$. In equilibrium, the fields are distributed according to the following Landau-Ginzburg free energy
\begin{subequations}
\begin{align}
    \mathcal H[\phi]  &= \int d^3x \left[\frac{1}{2} \nabla \phi_a\cdot
\nabla \phi_a + U(\phi) 
+\frac{1}{4\chizero} n_{ab}n^{ab}  \right]
    \label{eq:Landauginzburg}\\
    \label{eq:potential}
U(\phi) &= \frac{1}{2} m_0^2 \, \phi^2  + \frac{\lambda}{4} (\phi\cdot\phi)^2 \,
.%\quad \mbox{with} \quad m_0^2 < 0 \, .
\end{align}
\end{subequations}
 The dynamics is governed by the following equations of motion (EoM)
\begin{subequations}
    \label{eq:eom}
\begin{align}
   \partial_t \phi_a  + \frac{1}{\chizero}\,n_{ab} \phi_b  
  &=\Gamma_0 \frac{\delta \mathcal H}{\delta \phi_a}+ \theta_a\, ,\label{eq:eom1}\\
   \partial_t n_{ab}  + \nabla \cdot (\nabla \phi_{[a} \phi_{b]}) 
             &= \sigma_0 \nabla^2\frac{\delta \mathcal H}{\delta n_{ab}}+ \partial_{i} \Xi_{ab}^i \label{eq:eom2}\, .
\end{align}
\end{subequations}
The left-hand side of \cref{eq:eom} has a Poisson bracket structure and encodes the ideal superfluid dynamics discussed below. 
The right-hand side encodes the dissipative dynamics with the relaxation 
coefficient $\Gamma_0$  and conductivity $\sigma_0$. Finally, the noise 
terms,  $\theta$ and $\Xi$,  reproduce the fluctuation-dissipation theorem.

At fixed coupling $\lambda$, the model has a critical mass $m_c^2$; for 
example, for $\lambda = 4$, $m_c^2 = -4.811$~\cite{Florio:2021jlx, 
Florio:2023kmy}. For $m_0^2 > m_c^2$, the $O(4)$ symmetry remains 
unbroken, and the mean value of the order parameter $\phi_a$ (the chiral 
condensate in QCD) vanishes. For $m_0^2 < m_c^2$, the $O(4)$ symmetry is 
spontaneously broken to $O(3)$, producing a condensate,  $\llangle \phi_a\rrangle \equiv \bsigma n_a$ (with $n_an_a=1$),  and three pion Goldstone modes, $\vec{\pi}(t,\x) = (\pi_1, \pi_2, \pi_3)$, describing the fluctuations of $\phi_a(t,\x)$ orthogonal to $n_a$ and the direction of $\phi_a(t,\x)$.  

In the broken phase, the magnitude of the order parameter $\phi_a$ is approximately constant,  and the left-hand side of \labelcref{eq:eom} describes the 
coupling between conserved charges and Goldstone modes. When the 
dissipative terms are neglected, the resulting 
coupled equations correspond to an ideal non-abelian superfluid, reflecting the 
breaking of $O(4)$ symmetry~\cite{Son:1999pa, 
Grossi:2020ezz}. The ideal superfluid theory depends on just two 
parameters: $f^2 = \phi_a \phi_a$ and the susceptibility $\chi_I$, both 
of which can be determined from Euclidean measurements in lattice QCD~\cite{Bazavov:2019www, Ce:2022dax, Krasniqi:2024kwm}. 

A linearized analysis of the superfluid equations (including dissipation) 
yields the pion dispersion relation, 
$\omega(k) = v k - i D_A k^2 / 2$, with $v^2 = f^2 / \chi_I$~\cite{Son:2002ci}. 
Thus, the propagation of long-wavelength pions is characterized by two 
timescales: a ``ballistic" timescale $1/(v k)$ and a ``diffusive" timescale 
$1/(D_A k^2)$.

Even near the critical point, these timescales remain parametrically 
separated for $k \xi \ll 1$. Matching the hydrodynamic pion effective field theory (EFT) to the 
underlying critical theory gives $v \sim \xi^{(2 - d)/2} \sim \xi / \tau_R$, 
where $\xi$ is the diverging correlation length and 
$\tau_R \sim \xi^{d/2}$ is the relaxation time of the order 
parameter~\cite{Son:2001ff}. Meanwhile, the diffusion constant scales as 
$D_A \sim \xi^2 / \tau_R$~\cite{Son:2002ci}.

Thus for large systems with $L\gg \xi$ and $k \sim 1/L$, there are two parametrically separated timescales:
\st
t_B \equiv \frac{L}{v} \sim \tau_R \left( \frac{L}{\xi} \right) \ , 
\label{eq:tB}
\stp
and
\st
t_D \equiv \frac{L^2}{D_A} \sim \tau_R \left( \frac{L}{\xi} \right)^2 \ , 
\label{eq:tD}
\stp
and the hierarchy $\tau_R \ll t_B \ll t_D$ holds near the critical point.
As we will see, when the system undergoes an instantaneous quench from the symmetric to the broken phase, the hierarchy of timescales leads to a strong enhancement of the infrared spectrum of the Goldstone modes (pions).

\begin{figure}
    \centering
    \includegraphics[width=1\linewidth]{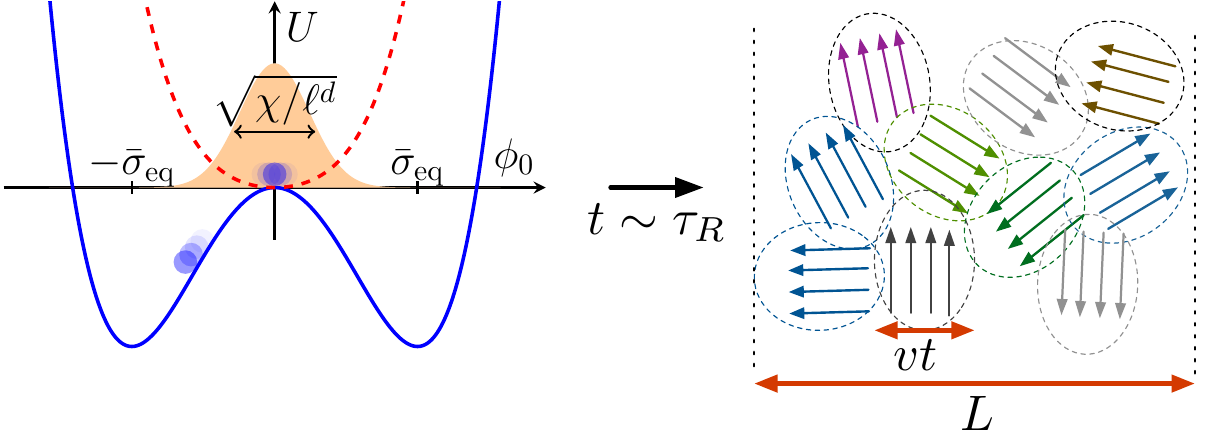}
    \caption{Illustration of instantaneous quenches in Model G. (Left) At every point in space, the effective potential of the theory suddenly changes from a convex potential to a double-well potential.
    After a time of order $\tau_R$, the field locally reaches a new minimum in field space, leading to many small randomly oriented domains shown in the right panel. (Right) The domains merge in a time $t_B \sim L/v$ with coarsening dynamics given by the superfluid limit of Model G (see text).}
    \label{fig:quenchsketch}
\end{figure}

\textit{Quenches and scaling analysis.}
We consider the setup shown in \cref{fig:quenchsketch}. The system begins in
thermal equilibrium at \( T > T_c \) (\( m_0^2 > m_c^2 \), red-dashed
potential) and is suddenly quenched into the broken phase at \( T < T_c \) (\(
m_0^2 < m_c^2 \), blue potential), with the quench tuned so that the
correlation length \( \xi \) remains the same before and after the quench. 
The post-quench field configuration is unstable.
In a region of size \( \ell \) (several correlation
lengths long), fluctuations of the locally averaged order parameter \(
\bar\phi_a \) are of order  \( \sim \sqrt{\chi/\ell^d} \), where \( \chi \sim
\xi^{2 - \eta} \) is the chiral susceptibility. These fluctuations seed the
instability, and the local order parameter finds  a new equilibrium value \( \bar\sigma_\text{eq}
\sim \xi^{-(d-2+\eta)/2} \) on a short timescale \(\sim  \tau_R \). At this
point, many small, randomly oriented domains have formed. On a longer timescale
\( t \sim L/v \), these domains begin to grow and merge. Since this growth time
\( t_B \sim L/v \) is much longer than the relaxation time \( \tau_R \), the
superfluid (hydrodynamic) limit of Model G provides an appropriate dynamical
description of the condensate growth. We will now elaborate on this picture.

As a proxy for the global condensate, we focus on ``magnetization"
$M_a(t)=\phi_a(t,\k=0)$, which is the zeroth mode of the Fourier transformed
order parameter  $\phi_a(t, \k) \equiv \frac{1}{V} \sum_{\bm x} e^{-i\k \cdot
{\bm x}} \phi_a(t, {\bm x})$. In addition, we consider the three pion fields
$\pi^\ell(t,\k)$, which are projections of $\phi_a(t,\k)$ orthogonal to $M_a$. In
particular, we study the non-equilibrium evolution of the following equal-time
two-point correlation functions
 \begin{align} 
 G_0(t) &\equiv \frac{V}{4}\sum_{a=1}^4 \llangle M_a(t) M_a(t)  \rrangle \ ,\label{eq:G0}\\
 G_{\pi\pi}(t,\k) &\equiv \frac{V}{3}\sum_{\ell=1}^3 \llangle \pi^\ell(t,\k) \pi^\ell(t,-\k)  \rrangle \ .\label{eq:Gpipi}
 \end{align}%
 Our companion paper details these definitions\citelongpaper.
 
At late times, when the quasiparticle approximation is valid,  the classical pion fields can be treated as fluctuations around a global condensate 
\st
(\sqrt{\chi_I}/\bsigma) \, \pi^{\ell}(t,{\bm x}) \simeq \sum_\k a_{\k}^{\ell} e^{-i\omega(k)t + i\k\cdot{\bm x}}/\sqrt{2\omega(k) V} + {\rm c.c.} 
\stp
The leading factor, $\sqrt{\chi_I}/\bsigma$, canonically normalizes the pion field to match the  quasi-particle description\citelongpaper. 
Then,  $\chi_I G_{\pi\pi}(t,k)/\bsigma^2 \simeq n_\pi(t,k)/\omega(k)$  estimates the soft-pion yield,  $n_{\pi}(t,k)= \sum_{\ell}\langle |a^\ell_\k|^2\rangle /3$.  
Below we will compare $G_{\pi\pi}$  to equilibrium, $G_{\pi\pi}/G_{\pi\pi}^{\rm eq} \simeq n_\pi(t, k)/n_{\pi}^{\rm eq}(k)$, finding a soft-pion enhancement post-quench.
In the future,  the stochastic field configurations at late times can be precisely matched onto a Boltzmann description~\cite{Grossi:2020ezz}.

\begin{figure}
    \centering
    \includegraphics[width=0.85\linewidth]{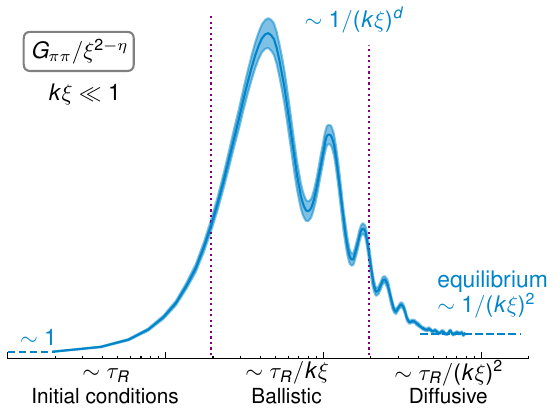}
    \caption{
        Schematic time evolution of the pion-pion correlator $G_{\pi\pi}(t,k)$ at small 
        momenta $k\xi\ll 1$,  following a symmetric quench at $t=0$ from the
        restored to the broken phase in three dimensions ($d=3$). The system
        first relaxes locally for $t\sim\tau_R$, seeding local condensate
        domains and  setting the initial conditions for correlator growth. Pions
        of size 
$1/k\gg\xi$ then propagate ballistically over time $t\sim\tau_R/(k\xi) \sim
1/vk$,   and the condensate domains grow to size $\sim 1/k$.  This coarsening dynamics in $d=3$ enhances 
$G_{\pi\pi}$ by $\sim 1/k\xi$   over
equilibrium, $G_{\pi\pi}^{\rm eq}\sim \xi^{2-\eta}/(k\xi)^2$.
Finally, the pion correlator returns to equilibrium (the dashed line) via diffusion for $t\sim\tau_R/(k\xi)^2$.
}
    \label{fig:param_enhancement_sum}
\end{figure}

In \cref{fig:param_enhancement_sum}, we provide a qualitative summary 
of the non-equilibrium time evolution of $G_{\pi\pi}(t,k)$ based on 
the following scaling discussion of $G_0(t)$ and then $G_{\pi\pi}(t,k)$.
At early times, the order parameter forms  small domains in a time  $\sim \tau_R$.  Domain alignment and the accompanying magnetization growth take place over the ballistic timescale $t_B \gg \tau_R$, i.e.~ the time needed to propagate across the volume. 
Finally, when the magnetization equilibrates  $M_a\sim \bsigma \, n_a$,  $G_0(t)$ approaches $V \bsigma^2/4\sim \xi^{2-\eta} \left(L/\xi\right)^d$. 
Having identified the timescales and asymptotes,  we formulate a scaling form for the non-equilibrium evolution of the correlation function $G_0$ for intermediate times $\tau_R\ll t\sim t_B \ll t_D$:
\st
G_0(t,\xi,L) = \xi^{2-\eta} \left(L/\xi\right)^d\mathcal{F}(vt/L,\xi/L) \ . 
\stp
So far,  since $v\sim \xi/\tau_R$,  this is a simple rewrite of the usual
statement of the dynamical scaling hypothesis, where all times are scaled by
$\tau_R\propto\xi^{\zeta}$ and all lengths are scaled by $\xi$. Now, however,
for $L \rightarrow \infty$ and  $vt/L$ fixed, we expect $\mathcal{F}$ to be a
regular function as $\xi/L\to 0$ and the $\xi/L$ dependence can be dropped: 
\st
G_0(t,\xi,L) = \xi^{2-\eta} \left(L/\xi\right)^d\mathcal{F}(vt/L) \ . \label{eq:scaling_G0}
\stp
This is a scaling prediction for the non-equilibrium evolution of condensate growth.
In an overlap region between early and ballistic times $\tau_R\ll t \ll t_B$, $G_0$ must be independent
of system volume, so $\mathcal F(vt/L)\propto (vt/L)^d$ and the magnetization grows
as a power-law,   $G_0\propto \xi^{2-\eta}(t/\tau_R)^d$.

We have identified a significant time region $\tau_R \ll t \sim t_B \ll t_D$,  where the condensate grows significantly following a quench. Since the relevant times,  $t \gg \xi/v$,  are large compared to the relaxation time $\tau_R$, this is a  
hydrodynamic regime with locally broken symmetry.  
However, dissipation does not become important until late
times,  $t \sim t_D$. Therefore, the appropriate hydrodynamic  
description  is given by the ideal superfluid  
equations discussed above.  
In the future, these superfluid equations can be simulated directly to determine condensate growth.
Although similar to coarsening in other critical systems~\cite{tauber2014critical},
the linear growth of domain size in time uniquely reflects the superfluid
limit of Model~G.

A similar scaling analysis can be applied to the pion correlation function $G_{\pi\pi}(t,k)$ in
\labelcref{eq:Gpipi}. Consider a regime where $\xi \ll 1/k \sim L$. For times
$t\lesssim t_B$, all four components of $\phi_a(t, \k)$  are uncorrelated and should
behave similarly (see \cref{fig:quenchsketch}(b)). We therefore expect the pion
correlator to scale similarly to $G_0$,  though the numerical result depends on $kL$, $
G_{\pi\pi}(t,\xi,k,L) = \xi^{2-\eta} \left(L/\xi\right)^d\mathcal{F}(vk t,kL)$.
Here, in analogy with \labelcref{eq:scaling_G0},  the residual $k\xi\ll 1$
dependence has been dropped,  while keeping $vkt \sim 1$
fixed. In an overlap region,  $1/L \ll k \ll 1/\xi$,  the
correlation function must become volume independent and we expect that
$\mathcal{F}(vkt, kL)$ simplifies
to $\mathcal{F}(vkt)/(kL)^d$. Thus,  in infinite volume,  the time evolution of the soft pion-correlator takes a simple non-equilibrium scaling form 
\begin{equation}
    G_{\pi\pi}(t, \xi, k) = \xi^{2-\eta}\, (k \xi)^{-d} \mathcal F( v kt ) \ , \label{eq:Gpipik}   
\end{equation}
for the extended temporal range $\tau_R\ll t\sim t_B \ll t_D$.

The scaling arguments can also be extended to a small but finite pion mass $m$,  incorporating explicit symmetry breaking.
The scaling form becomes $
G_{\pi\pi}(t,\xi,k,L, m) = \xi^{2-\eta} \left(L/\xi\right)^d\mathcal{F}(v m t,kL, m L)$. Assuming that $\mathcal{F}$ is volume independent for momenta such that $1/L\ll k\ll m \ll 1/\xi$ immediately implies that 
\st 
G_{\pi\pi}(t,\xi,k, m) = \xi^{2-\eta} (m\xi)^{-d} \mathcal{F}(vm t) \ .\label{eq:Gpipim}   
\stp

The effects of the non-equilibrium dynamics are striking. Noting that the
late-time equilibrium pion correlator is $ G^{\mathrm{eq}}_{\pi\pi}(\xi, k)\sim
\xi^{2-\eta}/{((k\xi)^2+(m\xi)^2)}$~\cite{Grossi:2021gqi},
\cref{eq:Gpipik,eq:Gpipim} imply a parametric enhancement over the equilibrium values for $d=3$:
\st 
\frac{G_{\pi\pi}(t, k)}{G^{\mathrm{eq}}_{\pi\pi}(k)} \underset{t\sim 1/vk}{\sim} \begin{cases} 1/k\xi\,,& k\gg m  \\ 1/{m\xi}\,,& k\ll m \end{cases} .\label{eq:enhancement}
\stp 
This is the key result of our analysis. We predict that during sudden quenches to the broken phase, the infrared spectrum of Goldstones (pions) is parametrically enhanced relative to equilibrium for a significant period of time, $\tau_R \ll t \sim t_B \ll t_D$,  see \cref{fig:param_enhancement_sum}.

\begin{figure}
    \centering
    \includegraphics[width=0.85\linewidth]{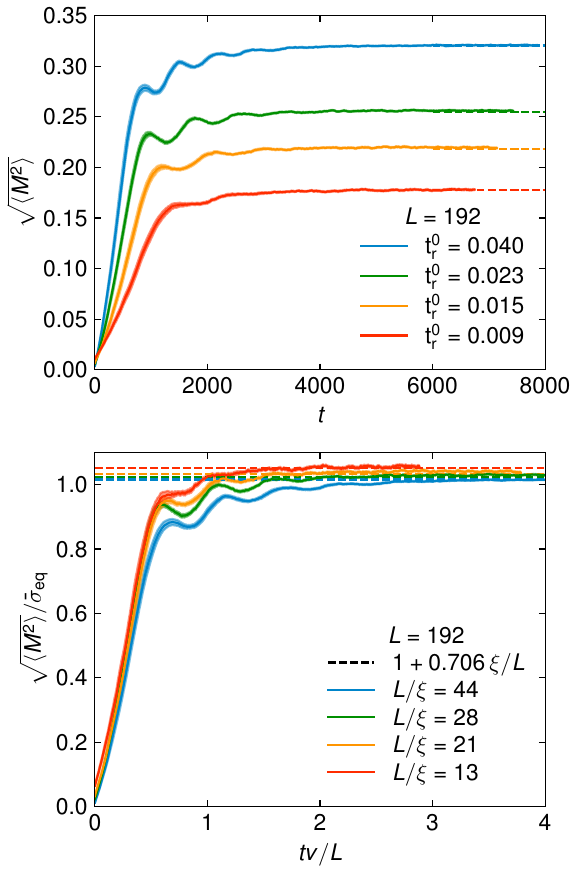}
    \caption{The chiral condensate (estimated by $\sqrt{\llangle M_a
    M_a\rrangle}$) versus time  following a symmetric quench,  $\tredini \rightarrow -\tredini$,   for several reduced temperatures,    $\tredini$. 
    The upper panel shows the raw data, while the lower panel shows the data with the $x$ and $y$ axes
    rescaled by $L/v$ and $\bsigma$ respectively. The data collapse (up to $\xi(\tred)/L$ corrections) confirms the non-equilibrium scaling predictions of
    \labelcref{eq:scaling_G0}. 
    }
    \label{fig:sigmafig}
\end{figure}

\begin{figure}
    \centering
    \includegraphics[width=0.85\linewidth]{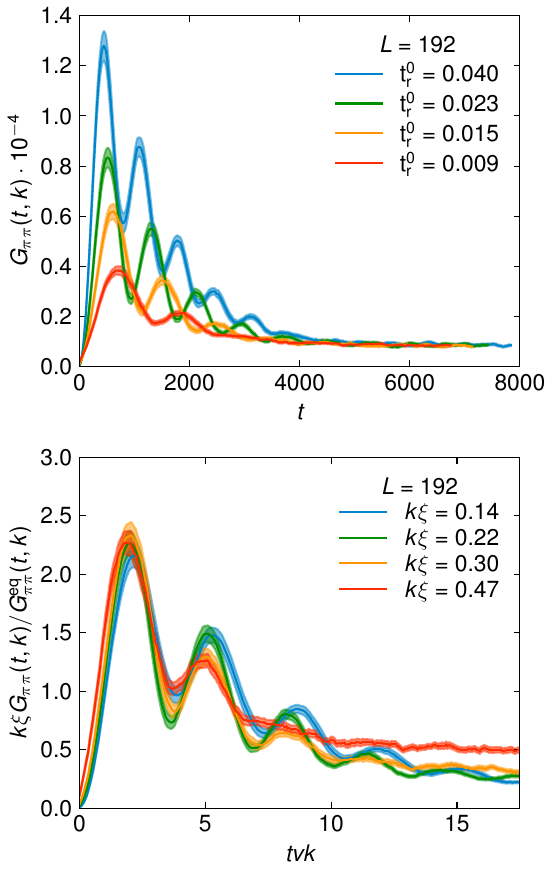}
    \caption{
Pion-pion correlator versus time at $k=2\pi/L$ following a symmetric quench, $\tredini\rightarrow -\tredini$,  for several initial reduced 
temperatures,   $\tredini$. The upper panel shows the raw data, while the 
lower panel shows  the data with the $x$ and $y$ axes rescaled by $vk$ 
and $k\xi/G_{\pi\pi}^{\rm eq}$ respectively. The data collapse  
at intermediate times in the lower panel supports the non-equilibrium 
scaling in \labelcref{eq:Gpipik} and implies that $G_{\pi\pi}(t,k)$ is 
parametrically enhanced by $\sim 1/k\xi$ at these times. At late times, 
the scaled correlators thermalize and the scaled correlators do not collapse. 
    }
    \label{fig:pionfig}
\end{figure}

\textit{Numerical simulations.} The scaling predictions can be tested by numerical simulations of \labelcref{eq:eom1,eq:eom2}.
The details are reported in the accompanying work\citelongpaper.
In \cref{fig:sigmafig,fig:pionfig}, we confirm the non-equilibrium time evolution from our dynamic scaling
analysis of the magnetization and pion correlation functions. \Cref{fig:sigmafig} shows the evolution of the \VEV, estimated from
$\sqrt{\langle M^2\rangle }$, following an instantaneous symmetric quench from
$\tredini$ to $-\tredini$. Here ${\sf t_r}\equiv(m_0^2-m^2_c)/|m_c^2|$ is the reduced
temperature and the correlation lengths in these quenches scale as 
$\xi\sim {|{\sf t_r}|^{-\nu}}$ with $\nu=0.7377(41)$~\cite{Engels:2014bra}.  Our 
scaling analysis is confirmed by the data collapse in the lower panel,  where time
is scaled by the ballistic time $t_B=L/v$ and the condensate is scaled by
$\bsigma$. Violations from scaling come from finite volume
corrections which distinguish $\sqrt{\langle M_a M_a\rangle}$ from
$\bsigma$~\cite{Engels:2014bra}\footnote{In practice, the finite-size
corrections to the magnetization can be parametrized as  $
\sqrt{\llangle M_a M_a\rrangle_{\rm eq}}  \approx \bsigma \left(1 + 0.706 \, \xi/L   \right) %\approx \bsigma (1 + 0.65 \, \xi/L} \, , 
$\citelongpaper.}.  

The non-equilibrium evolution of $G_{\pi\pi}(t,k)$ is shown in
\cref{fig:pionfig}. The top panel shows the raw data for quenches with
different initial $\tredini$ at fixed $k=2\pi/L$. Quench temperatures
starting farther from $T_c$ have smaller
$\xi$ and corresponding $k\xi$, and are more enhanced. After rescaling time with
$vk$ and the pion correlator with
$k\xi/G^{\rm eq}_{\pi\pi}$,  the data show good collapse
for $1\lesssim tvk\lesssim4$,  confirming the non-equilibrium scaling in \labelcref{eq:Gpipik} and implying a soft-pion enhancement. After the initial growth, $G_{\pi\pi}$ relaxes
slowly toward equilibrium on a diffusive timescale $t_D$,
and the non-equilibrium scaling no longer applies.

Much more can be said about the dynamics of instantaneous quenches. 
We refer the reader to the accompanying paper for
additional results on the evolution of fluctuations
parallel to $M_a$ (the sigma) and detailed mean-field
computations of $G_0(t)$ and $G_{\pi\pi}(t,k)$\citelongpaper.

\textit{Discussion.} The chiral phase transition in QCD spontaneously breaks
$O(4)\simeq SU(2)_L\times SU(2)_R$ symmetry below $T_c$. 
The dynamic critical behavior describing the transition between the restored and broken phases (which resembles the transition between a normal and superfluid) belongs to the universality class of Model G~\cite{hohenberg}.  Below $T_c$ the hydrodynamic
description in the chiral limit incorporates the Goldstone modes (the pions), forming a non-abelian superfluid~\cite{Son:1999pa}.

For the case of instantaneous quenches to the broken
phase, we use generic scaling arguments and explicit numerical simulations to
understand the non-equilibrium dynamics of two-point correlation functions and condensate growth
following a quench (see also the companion paper\citelongpaper). 
The key result of
our analysis is that at intermediate ballistic times, the infrared spectrum of the pion correlation function is parametrically enhanced over its
equilibrium expectation, see \cref{eq:enhancement}.

Preliminary work shows that this qualitative prediction is robust to explicit symmetry breaking, i.e.,  a small pion mass. 
We have also relaxed the quench setup by sweeping the temperature through the phase transition at a finite rate,  and we are finding similar results.
If real-world QCD lies outside the $O(4)$ scaling
window~\cite{Braun:2020ada,Braun:2023pew}, exhibiting only broken chiral
symmetry and a moderately light pseudo-Goldstone mode, the qualitative reasons
for the enhanced pion-yield remain intact. Indeed, as described above (see
\cref{fig:quenchsketch}),  the condensate growth and the pion enhancement are driven by the superfluid dynamics of the pion EFT, after being seeded with random small domains induced by the quench. Critical scaling determines the
temperature dependence of the EFT parameters and the initial domain size and is
essential for the data collapse seen in
\cref{fig:pionfig}. But the enhancement is more general, requiring only some
random seeds and the applicability of the hydrodynamic pion EFT (with some
parameters) at later times.

These results should be a
strong motivation for future heavy-ion experiments, such as ALICE3~\cite{ALICE:2022wwr}, to carefully measure the yield and correlations of
low momentum pions with $p_T$ of order the pion mass.
Of immediate relevance is the current measured yield of soft pions, which
rises above baseline hydrodynamic expectations for $p_T\lesssim400\,$MeV and
reaches a $\sim50\%$ enhancement near $p_T\sim m_\pi$
\cite{Lu:2024upk}.  A key difficulty for current hydrodynamic models, which currently do not include the dynamics of chiral symmetry breaking,  is the lack
of a soft momentum scale that would produce a deviation from equilibrium only at small
momentum.

Our simulations are an idealization of  heavy‑ion collisions and real-world QCD. Here we considered only instantaneous quenches in the chiral limit, noting that these simplifying features can be relaxed in future studies~\cite{superpaper6}.
Our work is part of a broader effort to understand the universal properties of non-equilibrium critical systems. However, the analysis naturally explains why low-momentum pions should be
preferentially enhanced relative to equilibrium, 
provided the expansion of a heavy ion collision through the chiral
transition is sufficiently rapid and the real-world pion mass is sufficiently light.  
An essential next step is to
implement the stochastic Model~$G$ dynamics 
within realistic hydrodynamic codes and with pion masses and chiral parameters motivated by Lattice QCD and other studies~\cite{HotQCD:2019xnw,Ce:2022dax,Braun:2020ada,Braun:2023pew}. 
This would quantify whether the non-equilibrium critical dynamics analyzed rigorously here (in a specific setup and limit of QCD) can account for the observed low‑$p_T$ excess in heavy ion collisions.

\section*{Acknowledgement}

   We would like to thank 
   Jürgen Berges, 
   Jean-Paul Blaizot,
   Kirill Boguslavski,
   Jannis Gebhard, 
   Alexander Kalweit,
   Govert Nijs,
   Thimo Preis, 
   Klaus Reygers, 
   Thomas Schäfer,
   Wilke van der Schee,
   Dam T. Son, 
   and Urs Wiedemann
   for helpful
     discussions.
     This work is supported by the DFG through Emmy Noether Programme (project number 545261797 (A.F.) and project number 496831614 (A.M.)) , through CRC 1225 ISOQUANT (project number 27381115) (A.M.). A.S. was supported by funding from Horizon Europe research and innovation programme under the Marie Skłodowska-Curie grant agreement No. 101103006 and the project N1-0245 of Slovenian Research Agency (ARIS). This work was supported by U.S. Department of Energy, Office of Science, Office of Nuclear Physics, Grant No. DE-SC0012704 (A.F.) and Grant
 No. DE-FG-02-88ER40388 (D.T.).
This research used resources of the National Energy Research Scientific Computing Center (NERSC), a Department of Energy Office of Science User Facility using NERSC award NP-ERCAP0028307. We thank ECT* for support at the workshop ``Attractors and thermalization in nuclear collisions and cold quantum gases," where we profited from helpful discussions.

The data and plotting routines used in the figures are available at~\cite{florio_2025_17600429}.
\bibliographystyle{utphys}
\bibliography{literature.bib}

\providecommand{\href}[2]{#2}\begingroup\raggedright\begin{thebibliography}{10}

\bibitem{Busza:2018rrf}
W.~Busza, K.~Rajagopal, and W.~van~der Schee, ``{Heavy Ion Collisions: The Big
  Picture, and the Big Questions},''
  \href{http://dx.doi.org/10.1146/annurev-nucl-101917-020852}{{\em Ann. Rev.
  Nucl. Part. Sci.} {\bfseries 68} (2018) 339--376},
  \href{http://arxiv.org/abs/1802.04801}{{\ttfamily arXiv:1802.04801
  [hep-ph]}}.

\bibitem{Pisarski:1983ms}
R.~D. Pisarski and F.~Wilczek, ``{Remarks on the Chiral Phase Transition in
  Chromodynamics},'' \href{http://dx.doi.org/10.1103/PhysRevD.29.338}{{\em
  Phys. Rev. D} {\bfseries 29} (1984) 338--341}.

\bibitem{Borsanyi:2018grb}
S.~Borsanyi, Z.~Fodor, J.~N. Guenther, S.~K. Katz, K.~K. Szabo, A.~Pasztor,
  I.~Portillo, and C.~Ratti, ``{Higher order fluctuations and correlations of
  conserved charges from lattice QCD},''
  \href{http://dx.doi.org/10.1007/JHEP10(2018)205}{{\em JHEP} {\bfseries 10}
  (2018) 205}, \href{http://arxiv.org/abs/1805.04445}{{\ttfamily
  arXiv:1805.04445 [hep-lat]}}.

\bibitem{HotQCD:2018pds}
{\bfseries HotQCD} Collaboration, A.~Bazavov {\em et~al.}, ``{Chiral crossover
  in QCD at zero and non-zero chemical potentials},''
  \href{http://dx.doi.org/10.1016/j.physletb.2019.05.013}{{\em Phys. Lett. B}
  {\bfseries 795} (2019) 15--21},
  \href{http://arxiv.org/abs/1812.08235}{{\ttfamily arXiv:1812.08235
  [hep-lat]}}.

\bibitem{HotQCD:2019xnw}
{\bfseries HotQCD} Collaboration, H.~T. Ding {\em et~al.}, ``{Chiral Phase
  Transition Temperature in ( 2+1 )-Flavor QCD},''
  \href{http://dx.doi.org/10.1103/PhysRevLett.123.062002}{{\em Phys. Rev.
  Lett.} {\bfseries 123} no.~6, (2019) 062002},
  \href{http://arxiv.org/abs/1903.04801}{{\ttfamily arXiv:1903.04801
  [hep-lat]}}.

\bibitem{Kaczmarek:2020sif}
O.~Kaczmarek, F.~Karsch, A.~Lahiri, L.~Mazur, and C.~Schmidt, ``{QCD phase
  transition in the chiral limit},'' {\em NIC Series} {\bfseries 50} (2020)
  193, \href{http://arxiv.org/abs/2003.07920}{{\ttfamily arXiv:2003.07920
  [hep-lat]}}.

\bibitem{Kotov:2021rah}
A.~Y. Kotov, M.~P. Lombardo, and A.~Trunin, ``{QCD transition at the physical
  point, and its scaling window from twisted mass Wilson fermions},''
  \href{http://dx.doi.org/10.1016/j.physletb.2021.136749}{{\em Phys. Lett. B}
  {\bfseries 823} (2021) 136749},
  \href{http://arxiv.org/abs/2105.09842}{{\ttfamily arXiv:2105.09842
  [hep-lat]}}.

\bibitem{Cuteri:2021ikv}
F.~Cuteri, O.~Philipsen, and A.~Sciarra, ``{On the order of the QCD chiral
  phase transition for different numbers of quark flavours},''
  \href{http://dx.doi.org/10.1007/JHEP11(2021)141}{{\em JHEP} {\bfseries 11}
  (2021) 141}, \href{http://arxiv.org/abs/2107.12739}{{\ttfamily
  arXiv:2107.12739 [hep-lat]}}.

\bibitem{Rajagopal:1992qz}
K.~Rajagopal and F.~Wilczek, ``{Static and dynamic critical phenomena at a
  second order QCD phase transition},''
  \href{http://dx.doi.org/10.1016/0550-3213(93)90502-G}{{\em Nucl. Phys. B}
  {\bfseries 399} (1993) 395--425},
  \href{http://arxiv.org/abs/hep-ph/9210253}{{\ttfamily arXiv:hep-ph/9210253}}.

\bibitem{Bjorken:1991xr}
J.~D. Bjorken, ``{A Full Acceptance Detector for SSC Physics at Low and
  Intermediate Mass Scales: An Expression of Interest to the SSC},''
  \href{http://dx.doi.org/10.1142/S0217751X92001885}{{\em Int. J. Mod. Phys. A}
  {\bfseries 7} (1992) 4189--4258}.

\bibitem{Blaizot:1992at}
J.-P. Blaizot and A.~Krzywicki, ``{Soft pion emission in high-energy heavy ion
  collisions},'' \href{http://dx.doi.org/10.1103/PhysRevD.46.246}{{\em Phys.
  Rev. D} {\bfseries 46} (1992) 246--251}.

\bibitem{Gavin:1993bs}
S.~Gavin, A.~Gocksch, and R.~D. Pisarski, ``{How to make large domains of
  disoriented chiral condensate},''
  \href{http://dx.doi.org/10.1103/PhysRevLett.72.2143}{{\em Phys. Rev. Lett.}
  {\bfseries 72} (1994) 2143--2146},
  \href{http://arxiv.org/abs/hep-ph/9310228}{{\ttfamily arXiv:hep-ph/9310228}}.

\bibitem{UA1:1982kzj}
{\bfseries UA1} Collaboration, G.~Arnison {\em et~al.}, ``{Search for Centauro
  Like Events at the {CERN} Proton Anti-proton Collider},''
  \href{http://dx.doi.org/10.1016/0370-2693(83)90790-6}{{\em Phys. Lett. B}
  {\bfseries 122} (1983) 189}.

\bibitem{UA5:1986zif}
{\bfseries UA5} Collaboration, G.~J. Alner {\em et~al.}, ``{An accelerator
  search at 900 GeV CM energy for the Centauro phenomenon},''
  \href{http://dx.doi.org/10.1016/0370-2693(86)91213-X}{{\em Phys. Lett. B}
  {\bfseries 180} (1986) 415--422}.

\bibitem{Augusto:1999vz}
C.~R.~A. Augusto, S.~L.~C. Barroso, V.~Kopenkin, M.~Moriya, C.~E. Navia, and
  E.~H. Shibuya, ``{Search for disoriented chiral condensate in cosmic gamma -
  hadron families},'' \href{http://dx.doi.org/10.1103/PhysRevD.59.054001}{{\em
  Phys. Rev. D} {\bfseries 59} (1999) 054001}.

\bibitem{MiniMax:1999mxv}
{\bfseries MiniMax} Collaboration, T.~C. Brooks {\em et~al.}, ``{A Search for
  disoriented chiral condensate at the Fermilab Tevatron},''
  \href{http://dx.doi.org/10.1103/PhysRevD.61.032003}{{\em Phys. Rev. D}
  {\bfseries 61} (2000) 032003},
  \href{http://arxiv.org/abs/hep-ex/9906026}{{\ttfamily arXiv:hep-ex/9906026}}.

\bibitem{ALICE:2013mez}
{\bfseries ALICE} Collaboration, B.~Abelev {\em et~al.}, ``{Centrality
  dependence of $\pi$, K, p production in Pb-Pb collisions at $\sqrt{s_{NN}}$ =
  2.76 TeV},'' \href{http://dx.doi.org/10.1103/PhysRevC.88.044910}{{\em Phys.
  Rev. C} {\bfseries 88} (2013) 044910},
  \href{http://arxiv.org/abs/1303.0737}{{\ttfamily arXiv:1303.0737 [hep-ex]}}.

\bibitem{ALICE:2019hno}
{\bfseries ALICE} Collaboration, S.~Acharya {\em et~al.}, ``{Production of
  charged pions, kaons, and (anti-)protons in Pb-Pb and inelastic $pp$
  collisions at $\sqrt {s_{NN}}$ = 5.02 TeV},''
  \href{http://dx.doi.org/10.1103/PhysRevC.101.044907}{{\em Phys. Rev. C}
  {\bfseries 101} no.~4, (2020) 044907},
  \href{http://arxiv.org/abs/1910.07678}{{\ttfamily arXiv:1910.07678
  [nucl-ex]}}.

\bibitem{Mazeliauskas:2019ifr}
A.~Mazeliauskas and V.~Vislavicius, ``{Temperature and fluid velocity on the
  freeze-out surface from $\pi$, $K$, $p$ spectra in pp, p-Pb and Pb-Pb
  collisions},'' \href{http://dx.doi.org/10.1103/PhysRevC.101.014910}{{\em
  Phys. Rev. C} {\bfseries 101} no.~1, (2020) 014910},
  \href{http://arxiv.org/abs/1907.11059}{{\ttfamily arXiv:1907.11059
  [hep-ph]}}.

\bibitem{Melo:2019mpn}
I.~Melo and B.~Tom\'a\v{s}ik, ``{Kinetic freeze-out in central heavy-ion
  collisions between 7.7 and 2760 GeV per nucleon pair},''
  \href{http://dx.doi.org/10.1088/1361-6471/ab5f03}{{\em J. Phys. G} {\bfseries
  47} no.~4, (2020) 045107}, \href{http://arxiv.org/abs/1908.03023}{{\ttfamily
  arXiv:1908.03023 [nucl-th]}}.

\bibitem{Devetak:2019lsk}
D.~Devetak, A.~Dubla, S.~Floerchinger, E.~Grossi, S.~Masciocchi,
  A.~Mazeliauskas, and I.~Selyuzhenkov, ``{Global fluid fits to identified
  particle transverse momentum spectra from heavy-ion collisions at the Large
  Hadron Collider},'' \href{http://dx.doi.org/10.1007/JHEP06(2020)044}{{\em
  JHEP} {\bfseries 06} (2020) 044},
  \href{http://arxiv.org/abs/1909.10485}{{\ttfamily arXiv:1909.10485
  [hep-ph]}}.

\bibitem{Nijs:2020roc}
G.~Nijs, W.~van~der Schee, U.~G\"ursoy, and R.~Snellings, ``{Bayesian analysis
  of heavy ion collisions with the heavy ion computational framework
  Trajectum},'' \href{http://dx.doi.org/10.1103/PhysRevC.103.054909}{{\em Phys.
  Rev. C} {\bfseries 103} no.~5, (2021) 054909},
  \href{http://arxiv.org/abs/2010.15134}{{\ttfamily arXiv:2010.15134
  [nucl-th]}}.

\bibitem{JETSCAPE:2020mzn}
{\bfseries JETSCAPE} Collaboration, D.~Everett {\em et~al.}, ``{Multisystem
  Bayesian constraints on the transport coefficients of QCD matter},''
  \href{http://dx.doi.org/10.1103/PhysRevC.103.054904}{{\em Phys. Rev. C}
  {\bfseries 103} no.~5, (2021) 054904},
  \href{http://arxiv.org/abs/2011.01430}{{\ttfamily arXiv:2011.01430
  [hep-ph]}}.

\bibitem{Lu:2024upk}
P.~Lu, R.~Kavak, A.~Dubla, S.~Masciocchi, and I.~Selyuzhenkov,
  ``{Quantification of the low-$p_{\rm T}$ pion excess in heavy-ion collisions
  at the LHC and top RHIC energy},''
  \href{http://arxiv.org/abs/2407.09207}{{\ttfamily arXiv:2407.09207
  [hep-ph]}}.

\bibitem{Schlichting:2019tbr}
S.~Schlichting, D.~Smith, and L.~von Smekal, ``{Spectral functions and critical
  dynamics of the O(4) model from classical-statistical lattice simulations},''
  \href{http://dx.doi.org/10.1016/j.nuclphysb.2019.114868}{{\em Nucl. Phys. B}
  {\bfseries 950} (2020) 114868},
  \href{http://arxiv.org/abs/1908.00912}{{\ttfamily arXiv:1908.00912
  [hep-lat]}}.

\bibitem{Grossi:2020ezz}
E.~Grossi, A.~Soloviev, D.~Teaney, and F.~Yan, ``{Transport and hydrodynamics
  in the chiral limit},''
  \href{http://dx.doi.org/10.1103/PhysRevD.102.014042}{{\em Phys. Rev. D}
  {\bfseries 102} no.~1, (2020) 014042},
  \href{http://arxiv.org/abs/2005.02885}{{\ttfamily arXiv:2005.02885
  [hep-th]}}.

\bibitem{Grossi:2021gqi}
E.~Grossi, A.~Soloviev, D.~Teaney, and F.~Yan, ``{Soft pions and transport near
  the chiral critical point},''
  \href{http://dx.doi.org/10.1103/PhysRevD.104.034025}{{\em Phys. Rev. D}
  {\bfseries 104} no.~3, (2021) 034025},
  \href{http://arxiv.org/abs/2101.10847}{{\ttfamily arXiv:2101.10847
  [nucl-th]}}.

\bibitem{Cao:2022csq}
X.~Cao, M.~Baggioli, H.~Liu, and D.~Li, ``{Pion dynamics in a soft-wall AdS-QCD
  model},'' \href{http://dx.doi.org/10.1007/JHEP12(2022)113}{{\em JHEP}
  {\bfseries 12} (2022) 113}, \href{http://arxiv.org/abs/2210.09088}{{\ttfamily
  arXiv:2210.09088 [hep-ph]}}.

\bibitem{Yao:2022fwm}
L.~H. Yao and U.~C. T\"auber, ``{Critical dynamics of the antiferromagnetic
  O(3) nonlinear sigma model with conserved magnetization},''
  \href{http://dx.doi.org/10.1103/PhysRevE.105.064128}{{\em Phys. Rev. E}
  {\bfseries 105} no.~6, (2022) 064128},
  \href{http://arxiv.org/abs/2204.11145}{{\ttfamily arXiv:2204.11145
  [cond-mat.stat-mech]}}.

\bibitem{Florio:2023kmy}
A.~Florio, E.~Grossi, and D.~Teaney, ``{Dynamics of the O(4) critical point in
  QCD: Critical pions and diffusion in model G},''
  \href{http://dx.doi.org/10.1103/PhysRevD.109.054037}{{\em Phys. Rev. D}
  {\bfseries 109} no.~5, (2024) 054037},
  \href{http://arxiv.org/abs/2306.06887}{{\ttfamily arXiv:2306.06887
  [hep-lat]}}.

\bibitem{Braun:2023pew}
J.~Braun {\em et~al.}, ``{Soft modes in hot QCD matter},''
  \href{http://dx.doi.org/10.1103/PhysRevD.111.094010}{{\em Phys. Rev. D}
  {\bfseries 111} no.~9, (2025) 094010},
  \href{http://arxiv.org/abs/2310.19853}{{\ttfamily arXiv:2310.19853
  [hep-ph]}}.

\bibitem{Roth:2024rbi}
J.~V. Roth, Y.~Ye, S.~Schlichting, and L.~von Smekal, ``{Dynamic critical
  behavior of the chiral phase transition from the real-time functional
  renormalization group},''
  \href{http://dx.doi.org/10.1007/JHEP01(2025)118}{{\em JHEP} {\bfseries 01}
  (2025) 118}, \href{http://arxiv.org/abs/2403.04573}{{\ttfamily
  arXiv:2403.04573 [hep-ph]}}.

\bibitem{Roth:2024hcu}
J.~V. Roth, Y.~Ye, S.~Schlichting, and L.~von Smekal, ``{Universal critical
  dynamics near the chiral phase transition and the QCD critical point},''  (9,
  2024) , \href{http://arxiv.org/abs/2409.14470}{{\ttfamily arXiv:2409.14470
  [hep-ph]}}.

\bibitem{Bray:1994zz}
A.~J. Bray, ``{Theory of phase-ordering kinetics},''
  \href{http://dx.doi.org/10.1080/00018739400101505}{{\em Adv. Phys.}
  {\bfseries 43} (1994) 357},
  \href{http://arxiv.org/abs/cond-mat/9501089}{{\ttfamily
  arXiv:cond-mat/9501089}}.

\bibitem{tauber2014critical}
U.~T{\"a}uber, {\em Critical Dynamics: A Field Theory Approach to Equilibrium
  and Non-Equilibrium Scaling Behavior}.
\newblock Cambridge University Press, 2014.
\newblock \url{https://books.google.com/books?id=hL4dAwAAQBAJ}.

\bibitem{Florio:2025lvu}
A.~Florio, E.~Grossi, A.~Mazeliauskas, A.~Soloviev, and D.~Teaney, ``{Quenching
  through the QCD chiral phase transition},''
  \href{http://dx.doi.org/10.1103/plfm-z5xx}{{\em Phys. Rev. D} {\bfseries 112}
  (2025) 114019}, \href{http://arxiv.org/abs/2504.03514}{{\ttfamily
  arXiv:2504.03514 [hep-lat]}}.
  \url{https://link.aps.org/doi/10.1103/plfm-z5xx}.

\bibitem{hohenberg}
P.~C. Hohenberg and B.~I. Halperin, ``Theory of dynamic critical phenomena,''
  \href{http://dx.doi.org/10.1103/RevModPhys.49.435}{{\em Rev. Mod. Phys.}
  {\bfseries 49} (Jul, 1977) 435--479}.
  \url{https://link.aps.org/doi/10.1103/RevModPhys.49.435}.

\bibitem{Florio:2021jlx}
A.~Florio, E.~Grossi, A.~Soloviev, and D.~Teaney, ``{Dynamics of the $O(4)$
  critical point in QCD},''
  \href{http://dx.doi.org/10.1103/PhysRevD.105.054512}{{\em Phys. Rev. D}
  {\bfseries 105} no.~5, (2022) 054512},
  \href{http://arxiv.org/abs/2111.03640}{{\ttfamily arXiv:2111.03640
  [hep-lat]}}.

\bibitem{Son:1999pa}
D.~T. Son, ``{Hydrodynamics of nuclear matter in the chiral limit},''
  \href{http://dx.doi.org/10.1103/PhysRevLett.84.3771}{{\em Phys. Rev. Lett.}
  {\bfseries 84} (2000) 3771--3774},
  \href{http://arxiv.org/abs/hep-ph/9912267}{{\ttfamily arXiv:hep-ph/9912267}}.

\bibitem{Bazavov:2019www}
A.~Bazavov {\em et~al.}, ``{Meson screening masses in (2+1)-flavor QCD},''
  \href{http://dx.doi.org/10.1103/PhysRevD.100.094510}{{\em Phys. Rev. D}
  {\bfseries 100} no.~9, (2019) 094510},
  \href{http://arxiv.org/abs/1908.09552}{{\ttfamily arXiv:1908.09552
  [hep-lat]}}.

\bibitem{Ce:2022dax}
M.~C\`e, T.~Harris, A.~Krasniqi, H.~B. Meyer, and C.~T\"or\"ok, ``{Aspects of
  chiral symmetry in QCD at T=128\,\,MeV},''
  \href{http://dx.doi.org/10.1103/PhysRevD.107.054509}{{\em Phys. Rev. D}
  {\bfseries 107} no.~5, (2023) 054509},
  \href{http://arxiv.org/abs/2211.15558}{{\ttfamily arXiv:2211.15558
  [hep-lat]}}.

\bibitem{Krasniqi:2024kwm}
A.~Krasniqi, M.~C\`e, R.~J. Hudspith, and H.~B. Meyer, ``{Hot QCD matter around
  the chiral crossover: A lattice study with O(a)-improved Wilson fermions},''
  \href{http://dx.doi.org/10.1103/PhysRevD.110.114506}{{\em Phys. Rev. D}
  {\bfseries 110} no.~11, (2024) 114506},
  \href{http://arxiv.org/abs/2407.01657}{{\ttfamily arXiv:2407.01657
  [hep-lat]}}.

\bibitem{Son:2002ci}
D.~T. Son and M.~A. Stephanov, ``{Real time pion propagation in finite
  temperature QCD},'' \href{http://dx.doi.org/10.1103/PhysRevD.66.076011}{{\em
  Phys. Rev. D} {\bfseries 66} (2002) 076011},
  \href{http://arxiv.org/abs/hep-ph/0204226}{{\ttfamily arXiv:hep-ph/0204226}}.

\bibitem{Son:2001ff}
D.~T. Son and M.~A. Stephanov, ``{Pion propagation near the QCD chiral phase
  transition},'' \href{http://dx.doi.org/10.1103/PhysRevLett.88.202302}{{\em
  Phys. Rev. Lett.} {\bfseries 88} (2002) 202302},
\href{http://arxiv.org/abs/hep-ph/0111100}{{\ttfamily arXiv:hep-ph/0111100
  [hep-ph]}}.
%%CITATION = HEP-PH/0111100;%%.

\bibitem{Engels:2014bra}
J.~Engels and F.~Karsch, ``{Finite size dependence of scaling functions of the
  three-dimensional O(4) model in an external field},''
  \href{http://dx.doi.org/10.1103/PhysRevD.90.014501}{{\em Phys. Rev. D}
  {\bfseries 90} no.~1, (2014) 014501},
  \href{http://arxiv.org/abs/1402.5302}{{\ttfamily arXiv:1402.5302 [hep-lat]}}.

\bibitem{Braun:2020ada}
J.~Braun, W.-j. Fu, J.~M. Pawlowski, F.~Rennecke, D.~Rosenbl{\"u}h, and S.~Yin,
  ``{Chiral susceptibility in ( 2+1 )-flavor QCD},''
  \href{http://dx.doi.org/10.1103/PhysRevD.102.056010}{{\em Phys. Rev. D}
  {\bfseries 102} no.~5, (2020) 056010},
  \href{http://arxiv.org/abs/2003.13112}{{\ttfamily arXiv:2003.13112
  [hep-ph]}}.

\bibitem{ALICE:2022wwr}
{\bfseries ALICE} Collaboration, ``{Letter of intent for ALICE 3: A
  next-generation heavy-ion experiment at the LHC},''
  \href{http://arxiv.org/abs/2211.02491}{{\ttfamily arXiv:2211.02491
  [physics.ins-det]}}.

\bibitem{superpaper6}
A.~Florio, E.~Grossi, A.~Mazeliauskas, A.~Soloviev, and D.~Teaney {\em in
  progress} .

\bibitem{florio_2025_17600429}
A.~Florio, E.~Grossi, A.~Mazeliauskas, A.~Soloviev, and D.~Teaney, ``Plotting
  routines and data files for 2504.03514 and 2504.03516,'' Nov., 2025.
\newblock \url{https://doi.org/10.5281/zenodo.17600429}.

\end{thebibliography}\endgroup

\end{document}